\documentclass[reprint,amssymb,amsmath,aps,showpacs,10 pt,floatfix,prb,longbibliography]{revtex4-2}
\def\TS2{$1T$-TaS$_2$}

\usepackage{graphicx}%
\usepackage{color}
\usepackage{epstopdf}
\usepackage{amssymb}
\usepackage{amsmath}
\usepackage{amsfonts}

\usepackage{xcolor}

\usepackage{color}
\usepackage[colorlinks,bookmarks=false,citecolor=darkblue,linkcolor=red,urlcolor=blue]{hyperref} 
\definecolor{darkred}{rgb}{0.7,0.0,0.0}

\definecolor{darkblue}{rgb}{0,0.02,0.45}

\definecolor{darkgreen}{rgb}{0.02,0.45,0.0}

\definecolor{violet}{rgb}{0.8,0.2,0.6}

\begin{document}

\preprint{Ellipsometry on 1T-TaS2}

\title{Interlayer coupling driven phase evolution in hyperbolic \TS2 }


\author{Achyut Tiwari}
\email{achyut.tiwari@pi1.uni-stuttgart.de}
\affiliation{1. Physikalisches Institut, Universit{\"a}t Stuttgart, Pfaffenwaldring 57, 70569 Stuttgart, Germany}

\author{Bruno Gompf}
\affiliation{1. Physikalisches Institut, Universit{\"a}t Stuttgart, Pfaffenwaldring 57, 70569 Stuttgart, Germany}
\author{Martin~Dressel}

\affiliation{1. Physikalisches Institut, Universit{\"a}t Stuttgart, Pfaffenwaldring 57, 70569 Stuttgart, Germany}

\begin{abstract}

 Understanding how microscopic interactions control macroscopic phase transitions is central to quantum materials, where charge-density waves (CDW), Mott-states and superconductivity often compete. In \TS2, this competition is tied to a sequence of CDW phases and a hysteretic metal–insulator transition, but details of the transition especially the role of interlayer coupling remain unresolved. In this work, spectroscopic ellipsometry is used to determine the uniaxial dielectric response of bulk \TS2\ from room temperature down to the commensurate insulating state. The room-temperature data reveal a natural type-II hyperbolic behavior in the visible range, with negative in-plane and positive out-of-plane permittivity.
Temperature-dependent ellipsometry combined with anisotropic Bruggeman effective medium analysis indicates that the metallic domains driving  percolation evolves from disc-like to needle-like, and that during heating, an additional intermediate phase shows up. Our results identify the transition in \TS2\ as a three-dimensional, interlayer-driven percolation process and establish this material as a natural, tunable hyperbolic medium.

\end{abstract}

\pacs{}
\maketitle

\section{Introduction}

 Layered transition metal dichalcogenides (TMDs) represent a versatile class of two-dimensional materials with remarkable electronic, optical, and structural properties that can be tuned by external parameters such as composition, stacking, strain, doping, and external fields \cite{CHOI2017}. These materials feature strong covalent intralayer bonding and weak van der Waals interlayer interactions, enabling facile exfoliation into monolayers and providing access to emergent quantum phenomena including charge density waves, superconductivity, and metal-insulator transitions \cite{novoselov2005two, geim2013van}. The ability to modulate their phase behavior and band structure on demand makes metallic TMDs especially attractive for next-generation nanoelectronic and photonic technologies \cite{fiori2014electronics}. Understanding and controlling the anisotropic electronic and optical response of metallic TMDs is therefore central to exploiting their functionality in emerging technologies \cite{Zhao21}.

Quasi-two dimensional \TS2\ is a prototypical member of the TMD family, notable for its complex electronic phase diagram featuring multiple charge density wave (CDW) phases and pressure-induced superconductivity \cite{Wilson1975, tosatti1976, sipos2008mott}. Upon cooling, the system undergoes a sequence of phase transitions: from an incommensurate metallic CDW phase (IC-CDW) above 350 K, to a nearly commensurate metallic CDW phase (NC-CDW) between 180 K and 350 K, and finally to a commensurate insulating CDW phase (C-CDW) below 180 K, which is characterized by a lattice distortion known as the star-of-David pattern \cite{ThomsonPRB94}. The metal-insulator transition (MIT) into the C-CDW phase is a first-order transition that displays a hysteresis, with an additional metastable or intermediate phase (often termed the T-CDW phase) appearing between 215 K and 280 K upon heating \cite{rossnagel2011origin}. These properties make \TS2\ an ideal platform for investigating the interplay of electron-electron and electron-phonon interactions, correlated electronic states, and the role of dimensionality in phase transitions.

Despite extensive studies, a complete understanding of the electronic phase evolution in \TS2\ remains elusive, in part because many experimental methods are only surface-sensitive. Techniques such as scanning tunneling microscopy/spectroscopy (STM/STS) \cite{Wang2024, Origin2024}, angle-resolved photoemission spectroscopy (ARPES) \cite{Wang2020}, and nano-infrared near-field imaging (nano-IR/SNOM) \cite{Frenzel2018} have provided valuable insights into CDW order, surface electronic structure, and the evolution of distinct insulating domains. However, these techniques primarily probe the topmost layers and are not sensitive to bulk behavior or the role of interlayer stacking in driving phase transitions. Theoretical and experimental studies suggest that the stacking configuration of CDW domains along the out-of-plane direction can significantly influence the electronic state, even switching the system between insulating and metallic phases depending on the interlayer stacking, underscoring the critical role of interlayer coupling \cite{Ritschel2018, Lee2023, Lee2019, ritschel2015orbital}. A clear picture linking the anisotropic bulk optical response to the microscopic phase evolution is still missing. Spectroscopic ellipsometry especially when combined with Bruggeman effective medium approximation (BEMA), offers a promising approach to bridge this gap by enabling bulk-sensitive, angle-resolved extraction of both in-plane and out-of-plane dielectric functions. Previous ellipsometric studies on thin gold films, VO$_2$ and organic conductors have demonstrated the effectiveness of this method for probing phase transitions and optical anisotropy, motivating its application to \TS2\ \cite{Hoevel2010, Voloshenko2018, Tiwari2025}.

\begin{figure}[ht]
	\centering
	\includegraphics[width=1\columnwidth]{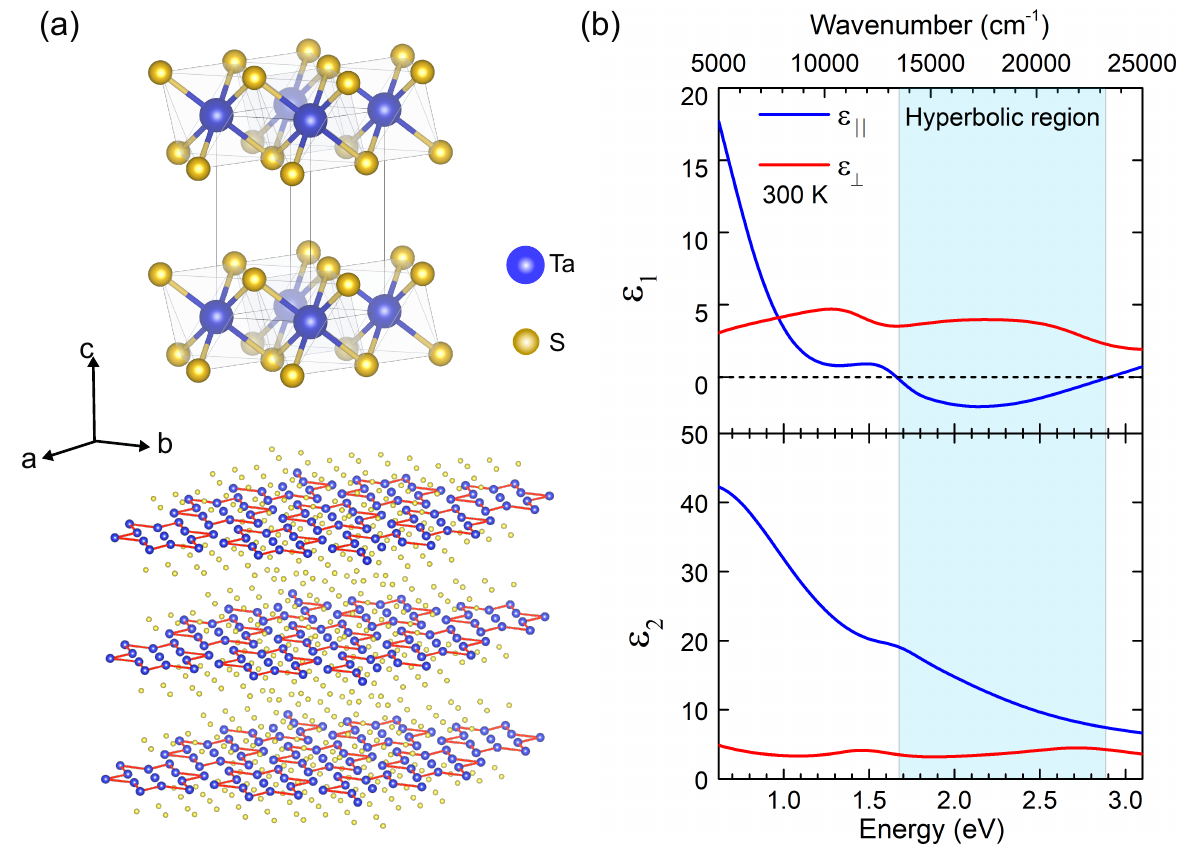}
	\caption{(a) Trigonal crystal structure of layered 1T-TaS2 (b) In-plane and out-of-plane real (upper panel) and imaginary (lower panel) parts of the complex dielectric function of \TS2\ {\it vs.} energy at room temperature. Opposite sign of real part of  the dielectric response in-plane and out-of-plane shows hyperbolic dispersion in the blue shaded energy region. }	
	\label{fig:fig1}
\end{figure}

In this paper, we employ spectroscopic ellipsometry at multiple angles of incidence to determine the uniaxial (in-plane {\it vs.} out-of-plane) complex dielectric functions, enabling the identification of hyperbolic dispersion in the NIR-vis range, marked by a negative in-plane and positive out-of-plane permittivity. 
Temperature-dependent ellipsometry discloses pronounced changes in both in-plane and out-of-plane dielectric functions across the phase transitions. By applying the anisotropic Bruggeman effective medium approximation ($a$BEMA), we further probe the microscopic details of the metal-insulator transition, revealing that metallic inclusions evolve anisotropically and elongate preferentially along the out-of-plane direction. These results highlight the pivotal role of interlayer coupling in governing the three-dimensional characteristics of phase transitions and offer important insights into the interplay between dimensionality and electronic phase evolution in layered correlated materials.

\begin{figure*}[t]
	\centering
	\includegraphics[width=1.75\columnwidth]{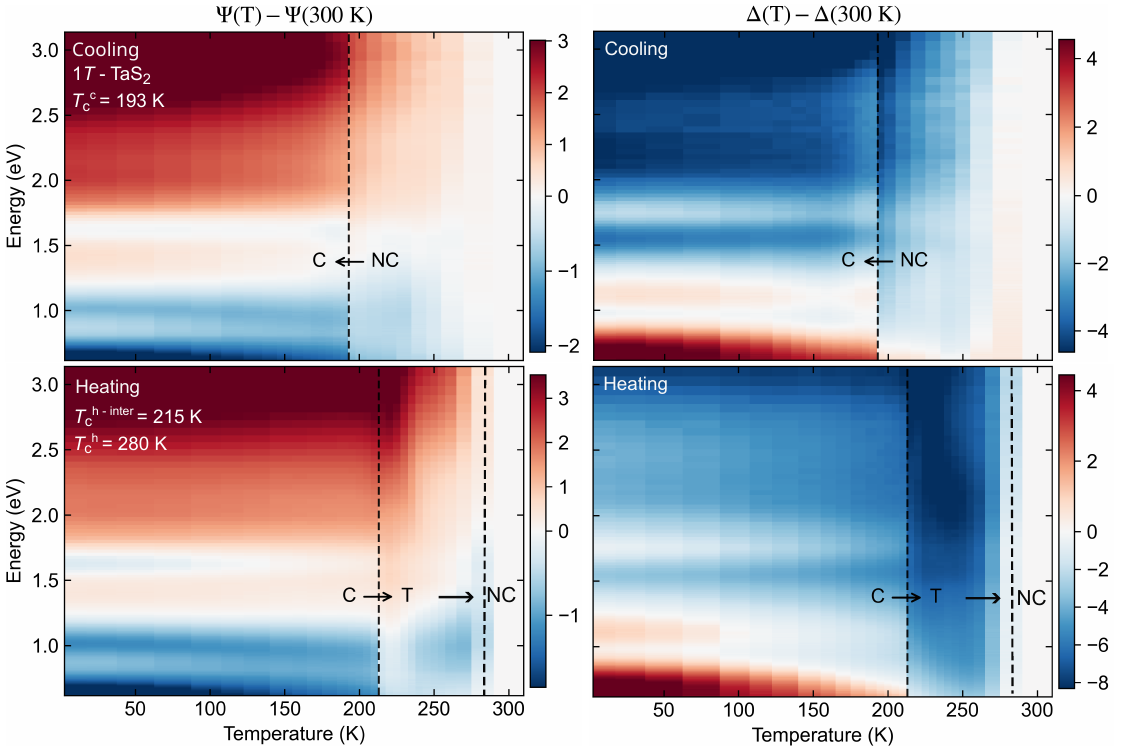}
	\caption{Temperature-dependent ellipsometric parameters $\mathrm{\Psi(T)-\Psi(300~K)}$ and $\mathrm{\Delta(T)-\Delta(300~K)}$ across metal-insulator for cooling (upper panels) and heating (lower panels). Vertical dashed lines shows the transition temperature from nearly commensurate (NC-metal) to commensurate (C-insulator) upon cooling. Upon heating, it displays the thermal hysteresis with an aditional intemediate phase denoted as triclinic (T) phase. }	
	\label{fig:fig2}
\end{figure*}

\section{Material and methods} 

High quality single crystal of \TS2\ (HQ graphene) was grown by  chemical vapor transport method. The sample was freshly cleaved immediately before measurement to obtain a flat, uncontaminated surface. The ellipsometric measurements on \TS2\ with lateral size of $\sim7~\mathrm{mm}\times5~\mathrm{mm}$ and a thickness of $\sim70~\mu\mathrm{m}$, has been performed under multiple angles of incidence (AOIs) at room temperature, while temperature-dependent ellipsometric measurements has been conducted only at a fixed AOI of ${70}^{o}$, due to restriction of the cryostat. Room-temperature spectroscopic ellipsometry was performed on a dual-rotating-compensator ellipsometer (RC2, J.A. Woollam Co., Inc.) in the wavelength range 400-1690~nm (0.73 - 3.1 eV). Multiple AOI scans were collected from $25^\circ$ to $70^\circ$ in $5^\circ$ steps. Temperature-dependent measurements were carried out from 300 K down to 10 K on a rotating-analyzer ellipsometer (VASE, J.A. Woollam Co., Inc.) coupled to a liquid-helium continuous-flow cryostat. Window effects from the cryostat were corrected using reference measurements on a standard silicon wafer. Data acquisition and model fitting were performed in CompleteEASE6~(J.~A.~Woollam~Co.,~Inc.).

\begin{figure}[t]
	\centering
	\includegraphics[width=1\columnwidth]{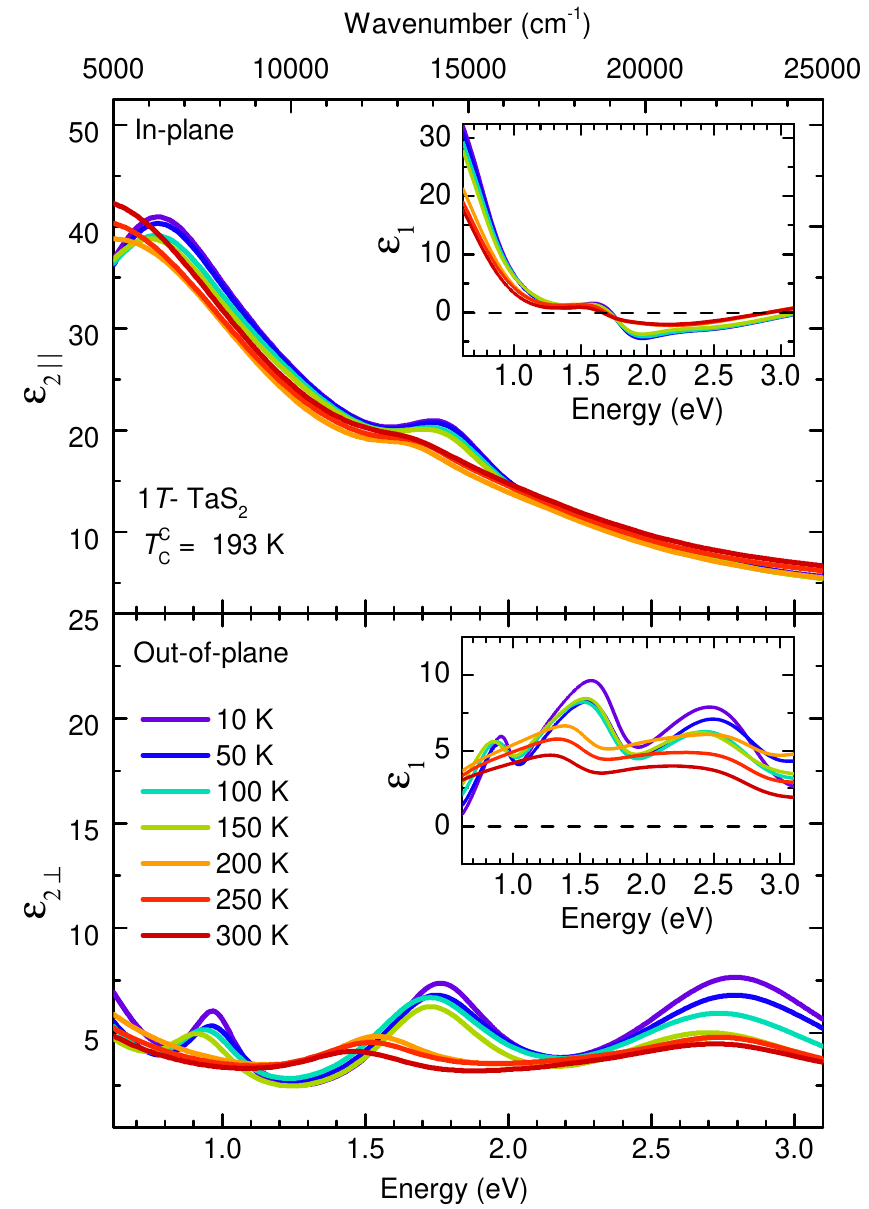}
	\caption{ Imaginary part of the complex dielectric function for in-plane (upper panel) and out-of-plane (lower panel) of \TS2\ {\it vs.} energy for different temperatures across metal-insulator transition. The insets show the corresponding real part of complex dielectric function {\it vs.} energy for different temperatures across metal-insulator transition.  The negative value of real part of in-plane dielectric constant down to low temperatures indicates presence of hyperbolicity in correlated insulating phase.}	
	\label{fig:fig3}
\end{figure}
 
\section{Results}

The complex uniaxial dielectric functions of \TS2\ at 300 K were obtained by analyzing the spectroscopic ellipsometric data at multiple AOIs using a uniaxial model that accounts for the anisotropy of layered \TS2\ with a trigonal crystal structure (Fig. \ref{fig:fig1}(a)).  A thin surface layer was included in the model using effective medium approximation i.e., the layer combining \TS2\ and void in a 50$\%$ mixture for the impact of surface roughness on the optical response. The in-plane response was modeled with a Drude term, one Tauc-Lorentz (2.05 eV) and three Lorentz oscillators (0.9 eV, 1.68 eV and 2.09~eV) to represent the free-carrier and interband configurations, in accord with previous reports \cite{Lin25}. The out-of-plane component used a Drude term and one Tauc-Lorentz (2.46~eV) and two Lorentz oscillators (1.5~eV and 2.77~eV).  The model shows excellent agreement with the experimental data (see Fig. S2 in the supplement material(SM)). Figure \ref{fig:fig1}(b) presents the real (upper panel) and imaginary (lower panel) parts of the dielectric function at room temperature, revealing significant anisotropy between in-plane and out-of-plane dielectric responses. The real parts of the in-plane and out-of-plane dielectric functions have opposite signs (the shaded region in blue in Fig.\ref{fig:fig1}(b)), indicating type-II hyperbolic dispersion ($\varepsilon_{1\parallel}$.$\varepsilon_{1\perp}$ $<$ 0) over a broad energy range, 1.67eV - 2.88~eV. Hyperbolicity over such a broad range in a natural van der Waals material, as theoretically predicted \cite{Gjerding2017}, makes \TS2\ promising for technological applications in the visible and near-infrared.

Figure \ref{fig:fig2} presents the differences in ellipsometric parameters, $\mathrm{\Psi(T)-\Psi(300~K)}$ and $\mathrm{\Delta(T)-\Delta(300~K)}$, for cooling (upper panel) and heating (lower panel). An abrupt change in both parameters, including a peak splitting near $\sim1$ eV, occurs at the transition temperature $T_c$~=~193 K, consistent with the first-order MIT reported for \TS2\ and its known hysteresis window \cite{Phonon2002}. The heating cycle exhibits a known broad temperature hysteresis and less discussed additional spectral rearrangements relative to cooling, indicative of distinct microscopic pathways on the cooling and heating, discussed in detail in the next section. 

The temperature-dependent measurements were analyzed in three steps. First, the uniaxial model set at room temmperature was used for the high-temperature metallic phase. For the insulating phase at 100 K, the same model was adopted but the Drude term was removed and two additional Lorentz oscillators were added. These two fits define the dielectric functions of the metallic and insulating end states. Since the transition is first order, and therefore could be described by the percolation transition. In the next step, all the temperatures in between were analyzed using anisotropic BEMA ($a$BEMA), with the metallic (230~K) and insulating (100 K) dielectric functions as two constituents. In the final step, it turned out that the heating cycle cannot be described accurately by a two-component mixture in the 225–280 K range, and an additional intermediate phase had to be introduced.

Figure \ref{fig:fig3} displays the temperature evolution of the in-plane and out-of-plane dielectric functions obtained from this analysis. In the insulating phase, additional Lorentz oscillators (in-plane: 1.26 eV and 0.58 eV; out-of-plane 0.97 eV and 0.55 eV) capture the opening of a low-energy gap and associated spectral weight transfer into interband features (see section S3 and S4 in SM for details), consistent with prior infrared work on the in-plane response of \TS2\ \cite{Lin25, Phonon2002}. We further provide the out-of-plane dielectric response with temperature, which is not reported in the literature, but is important to understand the role of interlayer interaction in the phase transition.  The temperature-dependent dielectric functions show substantial changes in both in-plane and out-of-plane components at $T_c$ = 193 K, indicating three-dimensional electronic phase transition with a strong interlayer interaction. Moreover, $\mathrm{Re}[\varepsilon_{\parallel}]$ remains negative over a wide low-energy window even in the insulating state, consistent with persistent plasmonic-like (delocalized) excitations coexisting with interband transitions reported recently by resonant inelastic X-ray scattering (RIXS) and band-theory analysis for insulating \TS2 \cite{Jia23}.

To further investigate the microscopic details of the transition, the optical response was analyzed using the anisotropic Bruggeman effective medium approximation, which treats the constituent phases symmetrically and is therefore suitable for interconnected or percolative mixtures of metallic and insulating domains; this contrasts with the Maxwell–Garnett approach, which assumes dilute inclusions in a host \cite{BEMA1935, MGEMA1904, Stroud1975, Choy}.

\begin{figure*}[t]
	\centering
	\includegraphics[width=1.85\columnwidth]{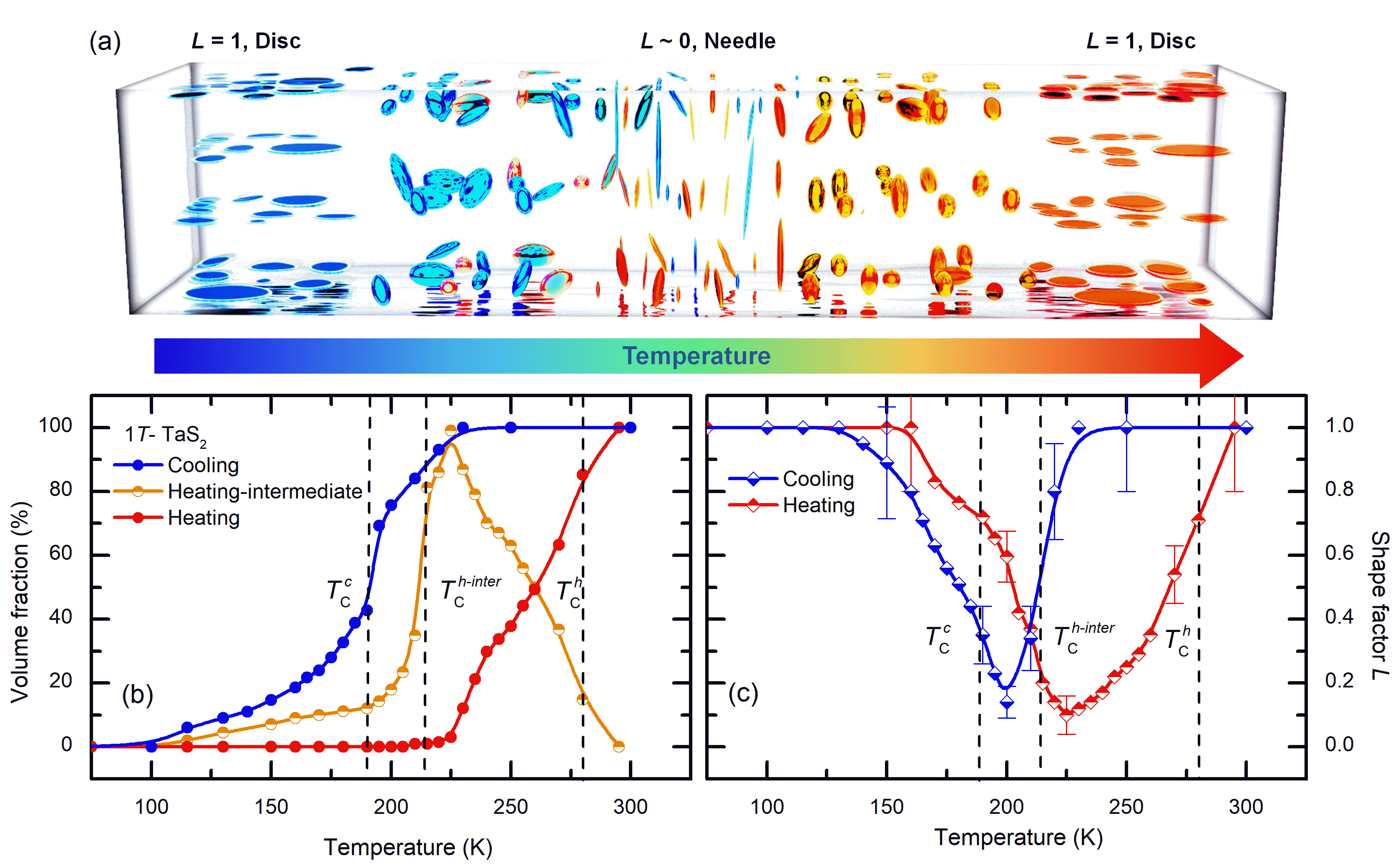}
	\caption{(a) Schematic illustration how the shape of the inclusions varies with temperature across the metal-insulator transition. (b) Volume fraction of metallic inclusions within insulating matrix extracted by using $a$BEMA together with associated (c) Shape factor across metal-insulator tansition upon cooling and heating. The error bars show the experimental uncertainty. The dashed vertical lines indicates the transition temperature.   }	
	\label{fig:fig4}
\end{figure*}

In the $a$BEMA for two-phase composites, the effective dielectric function $\varepsilon_{\rm eff}$ of a mixture of metallic and insulating domains with dielectric functions $\varepsilon_m$, and  $\varepsilon_d$, and volume fractions $f_m$, and  $f_d$ = 1 – $f_m$, is obtained by the following equation \cite{Taylor}:

\begin{equation}
 f_m\frac{\varepsilon_m-\varepsilon_{\rm eff}}{\varepsilon_{\rm eff}- L_i (\varepsilon_m-\varepsilon_{\rm eff})} +
 f_d\frac{\varepsilon_d-\varepsilon_{\rm eff}}{\varepsilon_{\rm eff}- L_i (\varepsilon_d-\varepsilon_{\rm eff})} = 0
 \label{eq:BEMA}
 \quad ,
\end{equation}
where $L_i$ is the direction-dependent ($i = x, y, z$) shape factor, which accounts for the anisotropy of the particle. Inclusions smaller than the wavelength can be approximated by ellipsoids with fixed revolution axis having a shape factor $L=(L_{\perp},L_{\parallel},L_{\parallel})$. For an isotropic case (sphere) one finds $L=(\frac{1}{3},\frac{1}{3},\frac{1}{3})$,
for a needle $(0,\frac{1}{2},\frac{1}{2})$, and for a disc $(1,0,0)$.

Here, $a$BEMA model was used, assuming that the optical response is fully characterized by its two constituents, the insulating and metallic phases. For the cooling, the dielectric function obtained for the low-temperature insulating phase (100 K) and high-temperature metallic phase (230 K) were kept fixed and only the metallic volume fraction and shape factor were allowed to vary in the $a$BEMA model. With these minimal parameters, the $a$BEMA model reproduces the spectra over a wide spectral range, with deviations near the transition temperature, as expected when properties change rapidly near percolation. Compared with an isotropic BEMA with a fixed shape factor, allowing anistropy in the shape factor  ($a$BEMA) yields improved agreement over the broad spectral range (see the FigS6 in SM).

For heating, the discrepancy between the two-component $a$BEMA model and the measurements is large, in the 225-280 K window. A three-component $a$BEMA model was therefore adopted, using dielectric functions obtained for the low-temperature insulating phase (100 K), the high-temperature metallic phase (300~K), and an intermediate phase ($\sim225$ K). This intermediate T-CDW phase is physically distinct from the high-temperature NC-CDW phase, exhibiting a stripe-like coexistence of insulating and metallic regions with comparable in-plane and out-of-plane conductivites (see the table S3 in SM). With only the phase volume fractions and a shape factor allowed to vary with temperature, the three-component $a$BEMA reproduces the spectra for heating, with deviation in the intermediate phase (see the FigS7 in SM). This behavior is consistent with reports of a broadened hysteresis and distinct intermediate textures on heating in \TS2 \cite{Geng2023}.

Figure \ref{fig:fig4} presents the temperature dependence of the metallic volume fraction (panel b) and shape factor (panel c) across transition. Upon cooling, the metallic volume fraction decreases gradually and then drops abruptly at the transition temperature, reaching negligible values by 100 K. Whereas on heating, a broad hysteresis is observed with the metallic inclusions remaining negligible up to $\sim215$ K, then increases and recovers the metallic state only above $\sim280$ K, with an intermediate phase in the window 215-280 K, in excellent agreement with previous reports \cite{Wang2020}.

\section{Discussion}

In order to comprehend how the extracted metallic volume fractions and shape factors reflect the underlying three-dimensional domain evolution, it is crucial to examine the behavior of the shape factor. Notably, the shape factor also exhibits hysteresis. During cooling, inclusions are disc-like ($L\sim1$) at high temperature, elongate toward needle-like ($L\sim0$) domains as it approaches to the transition, revert toward more isotropic values ($L\sim0.33$) near $T_c$, and finally recover a flatter, disc-like character in the insulating regime. During heating, the inclusions grow like flat-discs, where the metallic fraction is negligible and gradually become more ellipsoidal and needle-like near the transition and throughout the intermediate regime, and return to disc-like shapes at room temperature. This temperature-dependent variation of shape factor upon cooling and heating gives important insight into the intricate phase evolution driven by interlayer interaction.

Finally, the critical volume fraction and corresponding shape factor obtained from the $a$BEMA model were determined. A percolation threshold of $\mathit{f_m}$~$\sim43\%$ is obtained: above this value metallic NC-CDW domains form a connected network and below this threshold domains become isolated and the sample behaves insulating. This picture agrees well with nano-IR/SNOM measurements, where insulating C-CDW domains nucleate and expand until they surround and cut off the metallic regions, leaving only disconnected metallic islands \cite{Frenzel2018}. The temperature dependence of the shape factor shows that metallic domains are initially more disc-like and gradually evolve toward needle-like shapes as the insulating phase grows, suggesting that the metallic domains extend preferentially along the out-of-plane direction near the transition. The shape factor associated with the percolation threshold is approximately 0.35. In the $a$BEMA framework, the inclusion shape strongly influences the critical volume fraction at which percolation occurs \cite{Choy}. On heating, a three-component $a$BEMA that includes metallic, insulating and intermediate phases is required, which points to a different microscopic pathway for the C–NC transition. This intermediate phase is consistent with the triclinic CDW phase and associated electronic textures reported by STM, ARPES and temperature-dependent X-ray diffraction \cite{Geng2023, Wang2020, Burri2025}. In this regime, the fitted volume fractions and shape factors differ from those on cooling, indicating that the CDW domains break up and reconnect through distinct, often more filamentary pathways.

Our findings provide a bulk perspective that complements prior surface and local probes of \TS2. Our results are sensitive to the three-dimensional averaged electronic response, especially along the out-of-plane direction, which many techniques such as STM, ARPES cannot directly access. $a$BEMA results support a three-dimensional, percolative phase evolution in \TS2, where interlayer coupling and anisotropic domain morphology control how the metallic network forms and collapses across the hysteretic transition. The requirement of out-of-plane percolation for restoring metallic connectivity suggests that the low-temperature insulating state is better described as a band insulator stabilized by interlayer coupling, rather than a purely two-dimensional Mott insulator. However, the persistence of local in-plane metallicity and negative $\varepsilon_1$ deep in the insulating regime indicates that Mott-like correlations may still be relevant within isolated domains.

\section{Conclusion}
In conclusion, spectroscopic ellipsometry reveals the natural type-II hyperbolic optical response of \TS2\ at room temperature, and this behavior persists across its electronic phase transition. The temperature-dependent measurements provide direct access to both in-plane and out-of-plane dielectric functions, showing clear signatures of the metal-insulator transition and its associated hysteresis. The analysis using the anisotropic Bruggeman effective medium approximation ($a$BEMA) evidences that metallic domains evolve in an anisotropic manner and often extend along the out-of-plane direction as the transition proceeds. This microscopic evolution demonstrates that the phase transition in \TS2\ is inherently three-dimensional despite its layered structure, and that interlayer coupling plays a pivotal role in the phase evolution. Our results provide new insight into the microscopic phase evolution in \TS2\ and suggest that this material can serve as a platform for tunable optical and electronic devices based on its natural hyperbolic response.

\begin{acknowledgments}

The authors acknowledge the technical support from Gabriele Untereiner and discussion with R. Mathew Roy and Maxim Wenzel.
This work is supported by the Deutsche Forschungsgemeinschaft (DFG) under Grant No. DR228/63-1 and GO642/8-1.

\end{acknowledgments}

%

\pagebreak

\onecolumngrid

\newpage
\vspace*{-2.5cm}
\hspace*{-2.5cm} {
  \centering
  \includegraphics[width=1.2\textwidth, page=1]{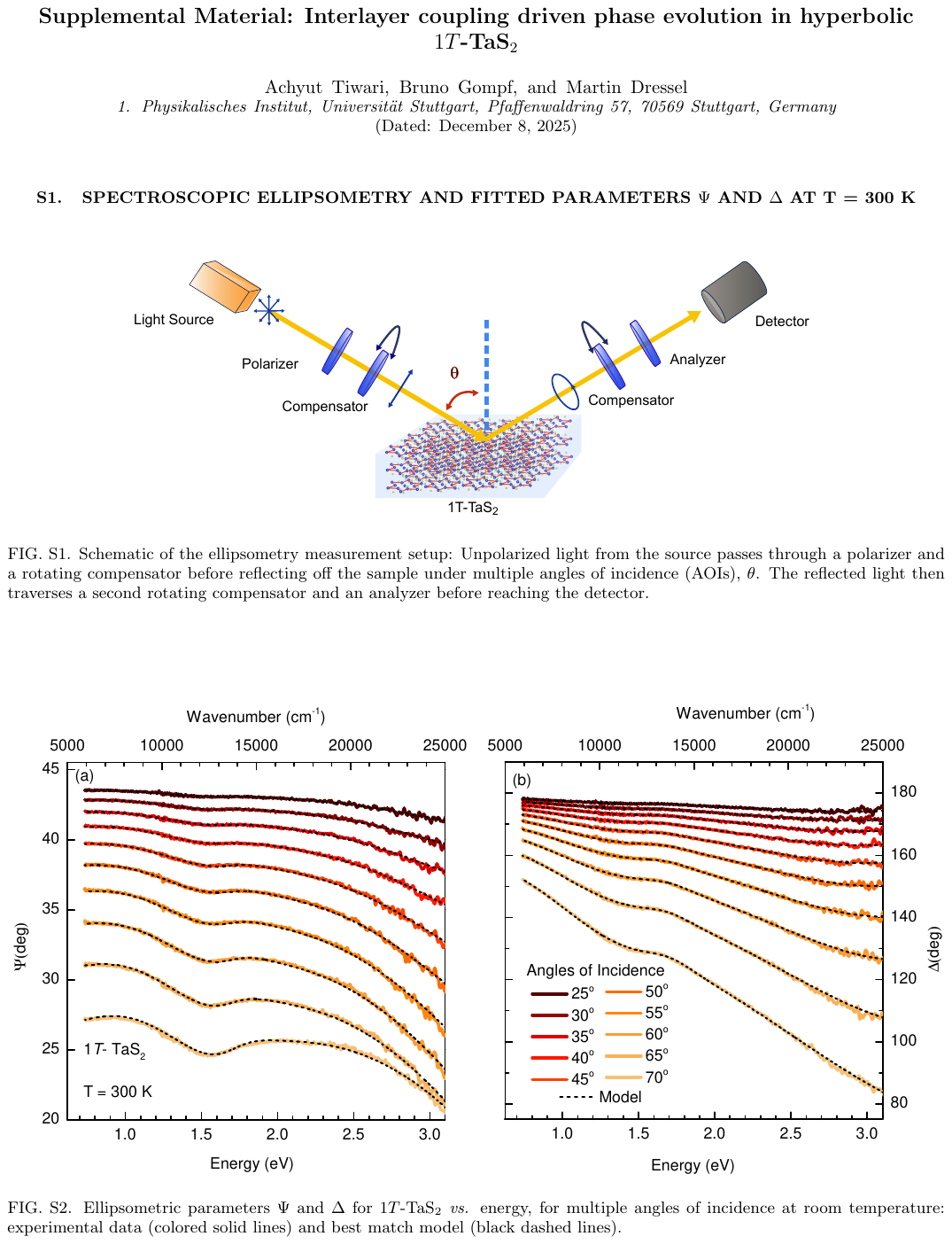} }
\vspace*{-2.5cm}
\hspace*{-2.5cm} {
  \centering
  \includegraphics[width=1.2\textwidth, page=2]{SM_TaS2.pdf} }

\vspace*{-2.5cm}
\hspace*{-2.5cm} {
  \centering
  \includegraphics[width=1.2\textwidth, page=3]{SM_TaS2.pdf} }

\vspace*{-2.5cm}
\hspace*{-2.5cm} {
  \centering
  \includegraphics[width=1.2\textwidth, page=4]{SM_TaS2.pdf} }

\vspace*{-2.5cm}
\hspace*{-2.5cm} {
  \centering
  \includegraphics[width=1.2\textwidth, page=5]{SM_TaS2.pdf} }

\vspace*{-2.5cm}
\hspace*{-2.5cm} {
  \centering
  \includegraphics[width=1.2\textwidth, page=6]{SM_TaS2.pdf} }

\vspace*{-2.75cm}
\hspace*{-2.5cm} {
  \centering
  \includegraphics[width=1.2\textwidth, page=7]{SM_TaS2.pdf} }
\end{document}